\documentclass[superscriptaddress,twocolumn]{revtex4}
\usepackage{graphicx,epsfig}
\usepackage{amsmath}
\usepackage {amssymb}
\usepackage[utf8]{inputenc}
\usepackage{float}

\usepackage[dvipsnames]{xcolor}
\RequirePackage[colorlinks,citecolor=blue,urlcolor=blue,linkcolor=blue]{hyperref}

\newcommand{\be}{\begin{eqnarray}}
\newcommand{\ee}{\end{eqnarray}}
\newcommand{\bea}{\begin{eqnarray}}

\newcommand{\eea}{\end{eqnarray}}

\def\k{\kappa}

\makeindex

\begin{document}

\title{Superradiant instability and charged scalar quasinormal modes for $2+1$-dimensional Coulomb-like AdS black holes from nonlinear electrodynamics}
\author{P. A. Gonz\'{a}lez}
\email{pablo.gonzalez@udp.cl} \affiliation{Facultad de
Ingenier\'{i}a y Ciencias, Universidad Diego Portales, Avenida Ej\'{e}rcito
Libertador 441, Casilla 298-V, Santiago, Chile.}
\author{\'Angel Rinc\'on}
\email{aerinconr@academicos.uta.cl} \affiliation{Sede Esmeralda, Universidad de Tarapac\'a,
Avda. Luis Emilio Recabarren 2477, Iquique, Chile.}
\author{Joel Saavedra}
\email{joel.saavedra@ucv.cl} \affiliation{Instituto de
F\'{i}sica, Pontificia Universidad Cat\'olica de Valpara\'{i}so,
Casilla 4950, Valpara\'{i}so, Chile.}
\author{Yerko V\'{a}squez}
\email{yvasquez@userena.cl}
\affiliation{Departamento de F\'{\i}sica, Facultad de Ciencias, Universidad de La Serena,\\
Avenida Cisternas 1200, La Serena, Chile.}

\date{\today}

\begin{abstract}
We study the propagation of charged scalar fields in the background of $2+1$-dimensional Coulomb-like AdS black holes, and we show that such propagation is unstable under Dirichlet boundary conditions. However,  all  the  unstable  modes  are  superradiant  and  all  the  stable  modes  are  non-superradiant, according with the superradiant condition. Mainly, we show that when the scalar field is charged the  quasinormal frecuencies (QNFs) are always complex, contrary to the uncharged case, where for small values of the black hole charge the complex QNFs are dominant, while that for bigger values of the black hole charge the purely imaginary QNFs are dominant. 
\end{abstract}

\maketitle


\tableofcontents


\section{Introduction}

Gravity at lower dimensions is usually considered a vibrant field of research. In particular, gravity at (2+1) dimensions has a few features which make such space-time quite interesting. To name a few, the absence of propagating degrees of freedom simplifies the things with respect to the (3+1)-dimensional counterpart. In addition, it should be pointed out that gravity in (2+1) dimensions is closed related to Chern-Simons theory    \cite{Witten:1988hc,Achucarro:1987vz,Witten:2007kt}.
Also, it is well-known that (2+1)-dimensional gravity shares certain properties of its higher dimensional analogs.
A few decades ago, a three-dimensional black hole was investigated in the presence of a  negative cosmological constant. This is precisely the case of the now well-known Bañados, Teitelboim, and Zanelli (BTZ hereafter) black hole \cite{Banados:1992wn,Banados:1992gq}.

Initially, the electrically charged BTZ black hole was also studied in Ref. \cite{Banados:1992wn} and subsequently, it was reviewed in \cite{Carlip:1995qv}. The charged and rotating black hole was presented in Ref. \cite{Martinez:1999qi}.
The inclusion of a charge $Q$ is parameterized via the usual Maxwell Lagrangian, and the corresponding lapse function shows a logarithmic dependence on the radial coordinate \cite{Martinez:1999qi}.
Notice that the (linear) Maxwell action is not invariant under conformal transformations of the metric. To have an action invariant under conformal transformations, we can take advantage of nonlinear electrodynamics as the source of the Einstein equation. Up to now, we have a vast series of papers where nonlinear electrodynamics is investigated. For instance see \cite{Balart:2014cga,Chabab:2020xwr,Garcia-Diaz:2017cpv} and references therein.
Some remarkable examples of nonlinear electrodynamics in the context of black hole physics in general relativity or alternative theories of gravity are: 
i) Born-Infeld \cite{Dey:2004yt,Cai:2004eh,Zou:2013owa,Boillat:1970gw,Fernando:2003tz,Jing:2010zp,deOliveira:1994in,Gullu:2010pc,Hendi:2016pvx,Hendi:2017oka}
ii) Power-Maxwell \cite{Hendi:2014mba,Hendi:2017mgb,Rincon:2017goj,Rincon:2018dsq,Panotopoulos:2018rjx,Rincon:2021hjj}, 
and
iii) regular charged black holes \cite{AyonBeato:1998ub,Ghosh:2018bxg}.
In particular, in Ref \cite{Cataldo:2000we} a non-trivial black hole solution in (2+1)-dimensional space-time in the presence of an Einstein-power-Maxwell electrodynamics was firstly derived satisfying the weak energy condition. It should be mentioned that the electric field has the Coulomb structure of a point charge in the Minkowski space-time, and the solutions describe charged (anti)–de Sitter space-times. Also, such a solution does not have a logarithmic contribution to the lapse function.

What is more, their thermodynamic properties were recently studied in Ref. \cite{Cataldo:2020cxm}. Be aware and notice that the invariance under conformal transformations is also recovered for arbitrary dimensions $n$ whether the Maxwell Lagrangian is raised to the $(n/4)^{th}$ power. Under such circumstances, a Coulomb-like electric field (in arbitrary dimensions) was obtained \cite{Hassaine:2007py}. Thus, these particular types of theories were extended to study the existence of hairy black hole solutions, as was pointed out in Ref~\cite{Cardenas:2014kaa}.  


Black holes, their stability, and also the behavior of the propagation of fields in black hole backgrounds have been considerably investigated for more than five decades.
In this respect, the seminal work performed by Regge and Wheeler \cite{Regge:1957a} was the starting point to dozens of papers regarding perturbations on black holes.
Quasinormal modes (QNMs hereafter) are distinctive frequencies with a non-vanishing imaginary part, which contain the information on how black holes evolve after the perturbation has been applied \cite{Konoplya:2011qq,Zerilli:1971wd}. 
Thus, after such perturbation, it responds by emitting gravitational waves. At this point, the importance of QNMs becomes evident: they have gained attention due to the recent detection of gravitational waves \cite{Abbott:2016blz}. Also, the QNMs have a recognizable relevance in the context of the correspondence AdS/CFT \cite{Maldacena:1997re, Aharony:1999ti,Horowitz:1999jd}. 
The quasinormal frequencies depend on: i) the type of geometry, and ii)  the type of the perturbation (i.e., scalar, vector, tensor, or fermionic), irrespectively of the initial conditions. 
Black hole perturbation theory \cite{Zerilli:1970se,Regge:1957td,Zerilli:1974ai} and QNMs become important during the ``ring down'' phase of a black hole merger. In such a stage is where a single distorted object is formed and where the geometry of space-time undergoes damped oscillations due to the emission of gravitational waves. Up to now, we have a significant collection of papers where the QNMs and  their corresponding quasinormal frequencies (QNFs) are computed. However, barely the most canonical black hole backgrounds have been investigated in detail. For an incomplete list of papers in this topic see \cite{Finazzo:2016psx,Gonzalez:2017shu,Gonzalez:2018xrq,Becar:2019hwk,Aragon:2020qdc, Aragon:2020tvq, Aragon:2020xtm, Aragon:2020teq, Fontana:2020syy,Rincon:2020pne,Rincon:2020iwy,Panotopoulos:2019gtn,Panotopoulos:2019qjk} and references therein. For an extensive review, see \cite{Konoplya:2011qq}.
Also, it has been found that most of the black holes are stable under certain types of perturbations (see \cite{Konoplya:2011qq} and references therein). In this respect, an astute way to investigate the stability of black holes is to perform the computation of the  QNMs and their QNFs \cite{Zerilli:1971wd,
Zerilli:1970se, Kokkotas:1999bd, Nollert:1999ji}. 

The QNMs for the BTZ background was studied in Ref. \cite{Cardoso:2001hn,Birmingham:2001pj}.
Also, it was shown that for scalar and fermionic fields the vanishing boundary conditions at infinity are automatically satisfied for the exact solutions, which implies a spectrum of QNFs without a decay rate for the extremal rotating BTZ black hole \cite{Crisostomo:2004hj}. Also, the Dirac quasinormal modes for rotating BTZ black holes with torsion were studied in Ref. \cite{Becar:2013qba}. The QNMs of the BTZ black hole for a conformal scalar field were studied in Ref. \cite{Chan:1996yk}. 
Also, the QNMs of the BTZ black hole surrounded by a conformal scalar field was analyzed in Ref. \cite{Konoplya:2004ik}, where it was estimated the shift in the quasinormal spectrum of the BTZ black hole stipulated by the backreaction of the Hawking radiation. See \cite{Gupta:2015uga, Gupta:2017lwk}, for the scalar, and fermionic quasinormal modes of the BTZ black hole in the presence of spacetime noncommutativity, respectively, and see  \cite{Becar:2014jia, Becar:2015kpa, Gonzalez:2015gla, Gonzalez:2017shu, Panotopoulos:2017hns, Gonzalez:2017zdz,Ciric:2017rnf, Rincon:2018sgd, Destounis:2018utr, Panotopoulos:2019qjk, Ciric:2019uab}, for other charged geometries.

In this work, we consider  $(2+1)$-dimensional Coulomb-like AdS black holes as background, and we study the stability of the propagation of a charged scalar field in order to study the effects of the scalar field charge on the propagation, by using the pseudospectral Chebyshev method \cite{Boyd}, which is an effective method to find high overtone modes \cite{Finazzo:2016psx,Gonzalez:2017shu,Gonzalez:2018xrq,Becar:2019hwk,Aragon:2020qdc, Aragon:2020tvq, Aragon:2020xtm, Aragon:2020teq, Fontana:2020syy}. Also, we study the superradiance phenomenon in this background, in order to analyze the stable and unstable modes, as we will show, all  the  unstable  modes  are  superradiant  and  all  the  stable  modes  are  non-superradiant, according to the superradiant condition.  
This work is organized as follows. In Sec.~\ref{background} we give a brief review of (2+1)-dimensional Coulomb-like AdS black holes. Then, in Sec.~\ref{QNM} we study the stability under Dirichlet boundary conditions. Then, in Sec.~\ref{superradiance} we analyze the superradiance phenomenon, and we calculate the QNFs of  charged scalar  perturbations numerically by using the pseudospectral Chebyshev method in Sec. ~\ref{QM}. Finally, we conclude in Sec.~\ref{conclusion}.

Along this manuscript, we will use geometrized units where $G =
c = 1$. We also use the most negative metric signature $(-, +, +)$

\section{Three-dimensional Coulomb-like AdS black holes}
\label{background}
We will start by considering the Einstein Hilbert action in  (2+1) dimensions in presence of a cosmological constant and  matter content described by nonlinear electrodynamics, minimally coupled with gravity. This action can be written as follows
 \begin{eqnarray} \label{action}
 S=\int d^{3}x\sqrt{-g}\left[\frac{1}{16 \pi}\left(R-2 \Lambda\right)+L(F)\right]~,
 \end{eqnarray}
where $R$ is the Ricci scalar, $\Lambda$ corresponds to the cosmological constant, and $L(F)$ represents the electromagnetic invariant Lagrangian,  and it is a nonlinear function of $F=\frac{1}{4} F_{\mu \nu} F^{\mu \nu}$. There are many black hole (BH) solutions for nonlinear electrodynamics \cite{Cai:2004eh, deOliveira:1994in, Hendi:2016pvx, Hendi:2017oka, Hendi:2014mba, Hendi:2017mgb}, and in particular for the case of power Maxwell invariant 
\begin{eqnarray} \label{action}
 S=\frac{1}{4}\int d^{3}x\sqrt{-g}\left( F_{\mu \nu} F^{\mu \nu}\right)^p~,
 \end{eqnarray}
the BH solution for conformal nonlinear electrodynamics and the traceless energy momentum tensor was found for the first time in \cite{Cataldo:2000we} for $p=\frac{3}{4}$
and later in Ref. \cite{Hassaine:2007py} it was obtained the BH solution for the  case $p=\frac{d}{4}$ where the conformal symmetry is manifestly. 
  Here we are focusing in Ref. \cite{Cataldo:2000we} because there are BH solutions for a vanishing trace energy-momentum tensor for the nonlinear electrodynamics under consideration in 2+1 gravity theory.  Here the electromagnetic nonlinear Lagrangian corresponds to $L(F)=C\left(  -F\right) ^{3/4}$,   
  and  the negative sign inside the Lagrangian guarantees purely real electric configurations.
 On the other hand, in $3+1$ dimensions for linear Maxwell electrodynamics, it is well known that the energy-momentum tensor is trace-free, and in this case, the solution for Maxwell equations is the standard very well known Coulomb solution. Now, for linear Maxwell theory minimally coupled to  $2 + 1$ dimensional gravity, the energy-momentum tensor has a not vanishing trace, and therefore  the electric field for a circularly symmetric static metric coupled to a Maxwell field is proportional to the inverse of $r$, i.e., $E_r \propto 1/r$. Hence the vector potential zero component $A_0$ is logarithmic, i.e., $A \propto \ln(r)$ and consequently blows up at $r = 0$, this solution correspond to the charged BTZ black hole \cite{Banados:1992wn}. Then, when we are considering nonlinear electrodynamics minimally coupled with $2+1$ gravity described by the electromagnetic Lagrangian $L(F)=C\left(  -F\right) ^{3/4}$, we can show the traceless energy-momentum tensor condition is satisfied and therefore, the resulting solution for the electric field is proportional to the inverse of $r^2$, surprisingly alike the Coulomb law for a point charge in 3 + 1 dimensions. Furthermore, the energy-momentum tensor satisfies the weak energy condition.
 
In our case, the circularly symmetric solution of this theory is given by the following metric %
\begin{eqnarray}
 ds^{2}=-f(r)dt^{2}+f^{-1}(r)dr^{2}+r^2 d \phi^2\,, \label{metricBH}
 \end{eqnarray}
and solving the Einstein-Maxwell equations under the condition of vanishing trace, we have
\begin{equation}
 T=T_{\mu \nu}g^{\mu \nu}=3L(F)-4FL,_F\,,
\end{equation}
which yields 
\begin{equation}
 L(F)=C|F|^{3/4}\,,
\end{equation}
 where $C$ is an integration constant. Because the magnetic field is vanished as a consequence of Einstein's equation, we get
 \begin{equation}\label{F}
 L(F)=C E^{3/2}\,,
\end{equation}
and from the Maxwell equation it follows that
 \begin{equation}
 E(r)=\left(\frac{Q^2}{6 \pi C}\right)^2\frac{1}{r^2}\,,
\end{equation}
where $Q$ is an integration constant
and finally, setting $C=\sqrt{|Q|}/6 \pi$, the electric field becomes
\begin{equation}
 E(r)=\frac{Q}{r^2}\,,
\end{equation}
a Coulomb-like electric field  but in $2+1$ dimensions.
Now, under the traceless condition, the components of Einstein equations $R_{tt}=-f^2 R_{rr}$, and $R_{\phi \phi}$ can be written as
\begin{eqnarray}
f_{,rr}+\frac{f_{,r}}{r}&=&-2 \Lambda +\frac{2Q^2}{3r^2}\,,\label{Rtt}\\
f_{,r}&=&-2 \Lambda-\frac{4Q^2}{3r^2}\,. \label{Rww}
\end{eqnarray}
It is easy to show Eq. (\ref{Rtt}) by virtue of the Maxwell equations. Therefore, the only remaining component of Einstein equations (\ref{Rww}) can be directly integrated, with the lapse function given by 
 \begin{equation}
 f(r)=-M-\Lambda r^2+\frac{4Q^2}{3r}\,, \label{f(r)}
 \end{equation}
 where $M$ is a constant related to the physical mass, and $Q$ is a constant related to the physical charge. In brief, we will return to this point later to discuss the physical significance of these constants.
Let us reinforce that this solution mimics those obtained in (3+1)-dimensional space-time for linear electrodynamics. The latter is an example of the electric field in light of the Einstein-power-Maxwell nonlinear electrodynamics, which, as we stated before, have been extensively studied (see \cite{Gurtug:2010dr,Liu:2012zza} and references therein).

The space-time is asymptotically de-Sitter space-time for $\Lambda > 0$, asymptotically flat for $\Lambda = 0$, and asymptotically anti de-Sitter for $\Lambda <0$. The roots of the lapse function are given by
\begin{align}
r_{h_{1}} &=\frac{h}{3 \Lambda}-\frac{M}{h}~,\,\,\label{rh}
\\
r_{h_{2}} &=-\frac{h}{6\Lambda}+\frac{M}{2h}+i\frac{\sqrt{3}}{2}\left(\frac{h}{3 \Lambda}+\frac{M}{h}\right)~,\,\,
\\
r_{h_{3}} &=-\frac{h}{6\Lambda}+\frac{M}{2h}-i\frac{\sqrt{3}}{2}\left(\frac{h}{3 \Lambda}+\frac{M}{h}\right),
\end{align}
being $h$ defined as follows
\begin{equation}
h=\left(\left(18q^2+3\sqrt{3\left(\frac{M^3}{\Lambda}+12Q^4\right)}\right)\Lambda^2\right)^{\frac{1}{3}}~.
\end{equation}

Here, we focus our study on the AdS case, where $M>0$. The solution shows different behaviors for the geometry depending on the value of the cosmological constant. There is a black hole solution with inner and outer horizons when $ 0>\Lambda >- \frac{M^3}{12Q^4}$,
there is one real and two complex solutions when $ \Lambda <- \frac{M^3}{12Q^4}$. Finally, when $\Lambda =- \frac{M^3}{12Q^4}$, the solution represents an extreme black hole. \\

\section{Charged scalar perturbations}
\label{QNM}

In order to study charged scalar perturbations in the background of  the metric (\ref{f(r)})
we consider the Klein-Gordon equation for charged scalar fields
\begin{equation}
\label{KGE}
\frac{1}{\sqrt{-g}}(\partial_{\mu}-iqA_{\mu})(\sqrt{-g}g^{\mu\nu}(\partial_{\nu}-iqA_{\nu})\psi)=m^2\psi ,
\end{equation}
plus suitable boundary conditions for a black hole geometry. In the above expression $m$ represents the mass and $q$ the charge of the scalar field $\psi$. Due to the circular symmetry, the Klein-Gordon equation can be written as 
\begin{equation}
\frac{d}{dr}\left(r f(r)\frac{dR}{dr}\right)+\left(\frac{r(\omega+qA_t(r))^2}{f(r)}-\frac{\kappa^2}{r}-m^{2}r \right) R(r)=0\,, \label{radial}
\end{equation}%
by means of the ansatz $\psi =e^{-i\omega t} e^{i\kappa \phi} R(r)$, where $\kappa=0, 1,2, \dots$, and $A_t=-\frac{Q}{r}$. Then, redefining $R(r)$ as $R(r)=\frac{F(r)}{\sqrt{r}}$, and by using the tortoise coordinate $r^*$ given by $dr^*=\frac{dr}{f(r)}$, the Klein-Gordon equation can be written as 
\begin{equation}
 \label{ggg}
 \frac{d^{2}F(r^*)}{dr^{*2}}-V_{eff}(r)F(r^*)=-\omega^{2}F(r^*)\,,
 \end{equation}
that corresponds to a one-dimensional Schr\"{o}dinger-like equation
 with an effective potential $V_{eff}(r)$, which is parametrically thought as $V_{eff}(r^*)$, and it is given by
 \begin{widetext}
  \begin{eqnarray}\label{pot}
 \nonumber V_{eff}(r)&=&\frac{f(r)}{r^2} \left(\kappa^2 + r\left( m^2r+ \frac{f^\prime(r)}{2}\right) - \frac{f(r)}{4}\right)-2\omega q A_t(r)-q^2A_t(r)^2~.\\
 \end{eqnarray}
 \end{widetext}
\newpage
%
Now, in order to study the stability of the propagation of scalar fields in the background of (2+1)-dimensional Coulomb-like AdS black hole, we follow the general argument given in Ref. \cite{Horowitz:1999jd}.
Thus, by replacing $\psi(r)=e^{i\omega r^{\ast}} F(r)$, in the Schr\"odinger-like equation \eqref{ggg} we obtain
\begin{equation}\label{KleinFink}
\frac{d}{dr}(f(r)\frac{d\psi(r)}{dr})-2i\omega \frac{d\psi(r)}{dr}-\frac{V_{eff}(r)}{f(r)}\psi(r)=0\,.
\end{equation}
Then, multiplying Eq. (\ref{KleinFink}) by $\psi^{\ast}$ and performing integrations by parts, where we have considered Dirichlet boundary condition for the scalar field at spatial infinity, it is possible to obtain the following expression
\begin{widetext}
\begin{equation}\label{relacion}
\int _{r_{+}}^{\infty}dr \left( f(r) \left|  \frac{d\psi}{dr}\right|^2+\frac{V_{eff}(r)|_{q=0}}{f(r)} \left| \psi \right| ^2  - \frac{q^2A_t(r)^2}{f(r)}\left| \psi \right| ^2\right)=-\frac{\left|\omega \right|^2 \left| \psi (r=r_{h})\right| ^2}{Im(\omega)}\,.
\end{equation}
\end{widetext}
So, notice that the sign outside the horizon of the expression $V_{eff}|_{q=0}-q^2A_t(r)^2$,  is crucial for stability.  For the neutral scalar perturbations the effective potential (\ref{pot}) is positive outside the horizon and then the left hand side of (\ref{relacion}) is strictly positive, which demand that $Im(\omega)<0$, and then the stability of the neutral scalar field under perturbations respecting Dirichlet boundary conditions is obeyed, which was pointed out in Ref. \cite{Aragon:2021ogo}. However, for charged scalar field, the integral can yield a negative value, therefore the stability is not guaranteed in this case. 

\section{Superradiant effect}
\label{superradiance}

\begin{figure}[H]
\begin{center}
\includegraphics[width=0.45\textwidth]{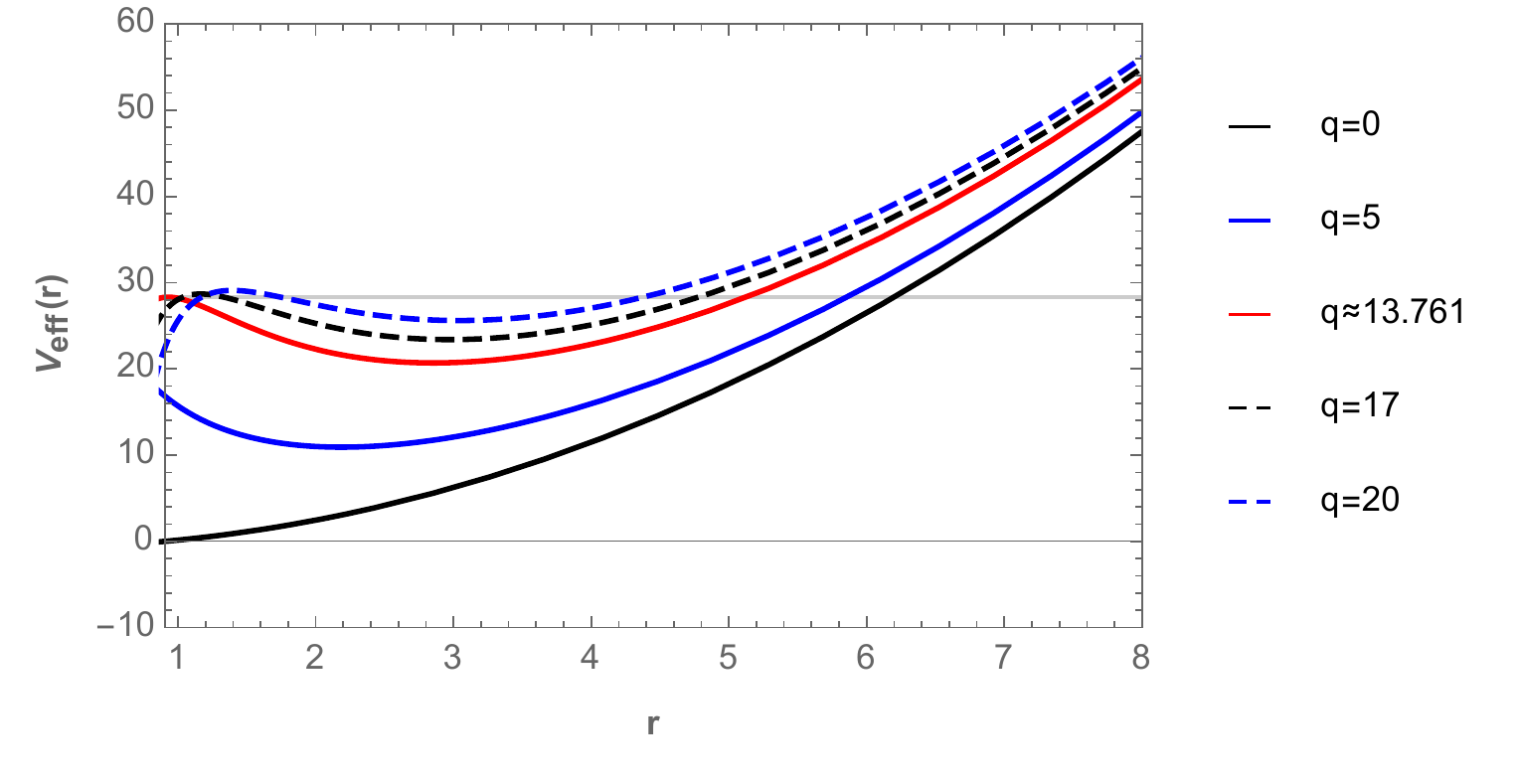}\\
\end{center}
\caption{The effective potential $V_{eff}(r)$ as a function of $r$, with $M=1$, $\Lambda=-1$, $Q=0.35$, $q=0,5,17,10$, $q_c\approx 13.761$, $m=0$, and $\kappa=0$. The horizontal line corresponds to $\omega^2\approx 28.3$.    }
\label{qcritical}
\end{figure}

The superradiant scattering of charged scalar field results in the extraction of both Coulomb energy and electric charge from the corresponding charged black hole. Then, this amplification of charged massive fields by charged black holes leads to instability as was shown for Reissner-Nordstr\"om space-time by Bekenstein \cite{Bekenstein}, for a recent review on superradiance see \cite{Brito:2015oca}. To find the conditions for superradiance amplification of scattering waves we will compute the greybody factor and the reflection coefficients. Then if the greybody factor is negative or the reflection coefficients is greater than 1 \cite{Benone:2015bst} then the scalar waves can  experiment a superradiant amplification by the black hole. Following  \cite{Harmark:2007jy} we will split the space-time in three regions and we will consider the low frequency limit, that is $\omega+qA_t(r_+) < < T_H$ and $(\omega+qA_t(r_+)) r_+ << 1$, and by simplicity we consider $\kappa=0$.
\begin{itemize}
\item
Region I: Corresponds to the region near the event horizon, which is defined by $r \approx r_+$. Here, the potential can be approximated as $V_{eff}(r) \approx - 2q \omega A_t(r_+)-q^2 A_t(r_+)^2$ or $V_{eff}(r)|_{q=0} << (\omega+qA_t)^2$.
In this region, the solution to the radial equation (\ref{radial}), is given by 
\begin{equation}
R(r)=A_Ie^{-i(\omega+qA_t(r_+))r^*}~,
\end{equation}
which slightly away the horizon yields
\begin{equation}
R(r)=A_I \left(1-\frac{i(\omega+qA_t(r_+))}{f'(r_+)}\log \left( \frac{r-r_+}{r_+}\right) \right)~.
\end{equation}

\item
Region II: Corresponds to the intermediate region, between the horizon and the asymptotic region. This region is defined by $V_{eff}(r)|_{q=0}>>(\omega+qA_t)^2$.\\

In this case the radial equation \eqref{radial} reduces to
\begin{equation}
\frac{d}{dr}\left(r f(r)\frac{dR}{dr}\right)=0~,
\end{equation}
whose solution is given by
\begin{equation}
R(r)=A_{II}+B_{II}G(r)~,
\end{equation}
where
\begin{equation}
G(r)=\int_{\infty}^r \frac{dr}{r f(r)}~.
\end{equation}
So, for $r\approx r_+$ we obtain
\begin{equation}
G(r)\approx\frac{1}{f'(r_+)r_+}\log\left(\frac{r-r_+}{r_+}\right)~.
\end{equation}
Matching this solution with the solution of region I, we obtain
\begin{equation}
A_{II}=A_{I},\,\,\, B_{II}=-i(\omega+qA_t(r_+))r_+A_{I}~.
\end{equation}

On the other hand, for $r>>r_+$
\begin{equation}
G(r)\approx\int_{\infty}^r \frac{dr}{(-\Lambda r^2)r}=\frac{1 }{2\Lambda r^2}.
\end{equation}
Therefore
\begin{equation}
R(r)=A_I\left(1-\frac{i(\omega+qA_t(r_+))r_+ }{2\Lambda r^2}\right),
\end{equation}
which will be matched with the asymptotic behavior.
\item
Region III: Corresponds to the asymptotic region, which is defined by $r>>r_+$.\\
In our particular case, it is sufficient to consider the leading term in the asymptotic behavior of the effective potential, i.e., 
\begin{equation}
V_{eff}(r)\approx  \frac{3 \Lambda^2 r^2}{4}~,
\end{equation}
Thus, the solution of the radial equation, in the asymptotic region, can be written as
\begin{equation}
R(r)= C_1+C_2/r^2~.
\end{equation}
Then, matching the solution of region II, for $r>>r_+$, with the solution of region III yields
\begin{equation}
C_1=A_I, \,\,\,  C_2= -i(\omega+qA_t(r_+))r_+A_I/(2\Lambda)~.
\end{equation}
\end{itemize}
After that, we will compute the fluxes utilizing the following expression
\begin{equation}
\mathcal{F}=\frac{\sqrt{-g}g^{rr}}{2i}\left( R^*\partial_r R-R\partial_r R^*\right)~.
\end{equation}
So, at the horizon we have
\begin{equation}
\mathcal{F}_{hor}\propto -(\omega+qA_t(r_+))r_+ |A_I|^2~.
\end{equation}
and at infinity
\begin{equation}
\mathcal{F}_{\infty}\propto -2\Lambda Im ( C_1C_2^* )~.
\end{equation}
In order to characterize the fluxes at the asymptotic region, it is convenient to split up the coefficients $C_{1}$ and $C_{2}$ in terms of the incoming and outgoing coefficients, $\hat{C}_2$ and $\hat{C}_{1}$, respectively. We define $C_1=\hat{C}_1+\hat{C}_2$ and $C_2=i(\hat{C}_2-\hat{C}_1)$. Therefore, the asymptotic incoming and outgoing fluxes are respectively given by

\begin{equation}
\mathcal{F}_{in\,\, \infty}\propto 2 \Lambda |\hat{C}_2|^2=2 \Lambda \frac{|A_I|^2}{4}\left(1-\frac{(\omega+qA_t(r_+))r_+}{2 \Lambda}\right)^2~,
\end{equation}

\begin{equation}
\mathcal{F}_{out\,\, \infty}\propto -2 \Lambda |\hat{C}_1|^2=-2 \Lambda \frac{|A_I|^2}{4}\left(1+\frac{(\omega+qA_t(r_+))r_+}{2 \Lambda}\right)^2~.
\end{equation}
Then, the reflection coefficient and the greybody factor \cite{Birmingham:1997rj,  Kim:1999un, Harmark:2007jy, Oh:2008tc, Kao:2009fh, Gonzalez:2010ht, Gonzalez:2010vv, Gonzalez:2011du} yields
\begin{equation}
\mathcal{R}= \left|\frac{\mathcal{F}_{out\,\, \infty}}{\mathcal{F}_{in\,\, \infty}} \right|=\left(\frac{1+\frac{(\omega+qA_t(r_+))r_+}{2\Lambda}}{1-\frac{(\omega+qA_t(r_+))r_+}{2\Lambda}} \right)^2~,
\end{equation}
\begin{equation}
\gamma(\omega)= \frac{\mathcal{F}_{hor}}{\mathcal{F}_{in\,\, \infty}}=\frac{-2(\omega+qA_t(r_+))r_+}{\Lambda\left(1-\frac{(\omega+qA_t(r_+))r_+}{\Lambda}\right)^2}~.
\end{equation}
Finally, based on the reflection coefficient (or from the greybody factor), it is possible to find the superradiant condition.
So the reflection coefficient is greater than 1 or alternatively the greybody factor is negative if $\omega+qA_t(r_+)<0$ or
\begin{equation}
    \omega<q\frac{Q}{r_+} \label{nbek}~,
\end{equation} 
 that coincides with the Bekenstein's superradiance condition \cite{Bekenstein}. In Fig. \ref{qcritical}, we show the behavior of the effective potential in order to visualize how it changes when the superradiance condition (\ref{nbek}) is satisfied or not. Note, for instance, that for  $\omega^2\approx 28.3$ and  $q>q_c$, with $q_c=\omega r_+/Q$  a potential well is possible, and there are bound states for charged scalar fields which allows to accumulate the energy to trigger the instability. However, for $q<q_c$ there are not bound states and the perturbation wave can be easily absorbed by the black hole and the corresponding background becomes stable under charged scalar perturbations.

In the next section, we will study the quasinormal modes and we will show that by comparing the real part of the dominant modes, with the superradiant condition of charged scalar fields \cite{Cardoso:2004hs, Cardoso:2006wa, Aliev:2008yk, Dias:2011at, Li:2012rx, Uchikata:2011zz, Cardoso:2013pza} suggests that all the unstable modes are superradiant unstable and consequently the scalar waves can experiment a superradiant amplification by the black hole, as in Ref. \cite{Gonzalez:2017shu}.\\

\section{Quasinormal modes}
\label{QM}

Now, in order to solve numerically the differential equation (\ref{radial}) we consider the pseudospectral Chebyshev method \cite{Boyd}, for an extensive review of numerical methods, see \cite{Konoplya:2011qq}. Firstly, it is convenient to perform the change of variable  $y=1-r_H/r$ in order to bound the value of the radial coordinate to the range $[0,1]$, and the radial equation (\ref{radial}) becomes
\begin{widetext}
\begin{equation} \label{r}
(1-y)^3 f(y) R''(y) +\left( (1-y)^3 f'(y)-(1-y)^2 f(y) \right) R'(y) + \left( \frac{(\omega+qA_t(y))^2 r_H^2}{f(y) (1-y)}- \kappa^2 (1-y) -\frac{m^2 r_H^2}{1-y} \right) R(y)=0\, ,
\end{equation}
\end{widetext}
where the prime means derivative with respect to the coordinate $y$. Now, the event horizon is located at $y=0$ and the spatial infinity at $y=1$. So, in order to propose an ansatz for the field, we analyze the behavior of the differential equation at the horizon and at infinity. In the neighborhood of the horizon the function $R(y)$ behaves as 
\begin{equation}
\label{horizon}
R(y)=C_1 e^{-\frac{i (\omega+qA_t(0)) r_H}{f'(0)} \ln{y}}+C_2 e^{\frac{i (\omega+qA_t(0)) r_H}{f'(0)} \ln{y}} \,,
\end{equation}
where: i) the first term represents an ingoing wave, and ii) the second represents an outgoing wave near the black hole horizon. So, imposing the requirement of only ingoing waves on the horizon, we fix $C_2=0$. On the other hand, at infinity the function $R(y)$ behaves as
\begin{equation}
R(y)= D_1 (1-y)^{1 + \sqrt{1 -\frac{ m^2}{\Lambda}}}+ D_2 (1-y)^{1 - \sqrt{1 -\frac{ m^2}{\Lambda}}} \,.
\end{equation}
So, imposing that the scalar field vanishes at infinity requires $D_2=0$. Therefore, an ansatz for $R(y)$ is $R(y) = (1-y)^{1 + \sqrt{1 -\frac{ m^2}{\Lambda}}}e^{-\frac{i (\omega+qA_t(0)) r_H}{f'(0)} \ln{y}} F(y)$, and by inserting this expression in Eq. (\ref{r}), it is possible to obtain a differential equation for the function $F(y)$. Now, to use the pseudospectral method, $F(y)$ must be expanded in a complete basis of functions $\{\varphi_i(y)\} $: $F(y)=\sum_{i=0}^{\infty} c_i \varphi_i(y)$, where $c_i$ are the coefficients of the expansion, and we choose the Chebyshev polynomials as the complete basis, which are defined by $T_j(x)= \cos (j \cos^{-1}x)$, where $j$ corresponds to the grade of the polynomial. The sum must be truncated until some $N$ value, therefore the function $F(y)$ can be approximated by  
\begin{equation}
F(y) \approx \sum_{i=0}^N c_i T_i (x)\,.
\end{equation}
Thus, the solution is assumed to be a finite linear combination of the Chebyshev polynomials,
that are well defined in the interval $x \in [-1,1]$. Due to $y \in [0,1]$, the coordinates $x$ and $y$ are related by $x=2y-1$.

Then, the interval $[0,1]$ is discretized at the Chebyshev collocation points $y_j$ by using the so-called Gauss-Lobatto grid, where
\begin{equation}
    y_j=\frac{1}{2}[1-\cos(\frac{j \pi}{N})]\,, \,\,\,\, j=0,1,...,N \,.
\end{equation}
The corresponding differential equation is then evaluated at each collocation point. So, a system of $N+1$ algebraic equations is obtained, which corresponds to a generalized eigenvalue problem and it can be solved numerically to obtain the QNMs spectrum, by employing the built-in Eigensystem[ ] procedure in Wolfram’s Mathematica \cite{WM}.\\

In this work, we use a value of $N$ into the interval [80-100] for the majority of the cases with an average running time in the range [80s-140s] which depends on the convergence of $\omega$ to the desired accuracy.  We will use an accuracy of eight decimal places. In addition, to ensure the accuracy of the results, the code was executed for several increasing values of $N$ stopping when the value of the QNF was unaltered. Also, the complete parameter space associated to the models is $M\geq 0$, $\Lambda<0$, and $\kappa= 0, 1, 2, ...$. Here, the regions of the parameter space explored is $M=1$, $\Lambda=-1$, and for the black hole charge we consider a discrete set of values in the interval [0, 0.52], due to in this region is guaranteed the existence of two positive real roots for the lapse function, and a discrete set of values of $\kappa$ in the interval [0, 30]. Also, for the scalar field mass we consider a discrete set of values in the interval [0, 0.3], and for the scalar field charge a discrete set of values in the interval [-0.09,15]. \\

\subsection{Massless charged scalar fields}

{\bf{Case $\kappa=0$}}. Is was shown that the QNMs for uncharged massless scalar field with $\kappa=0$ are purely imaginary \cite{Aragon:2021ogo}. So, in order to show the behavior of the QNMs for massless charged scalar fields, with $\kappa=0$, we fix the black hole mass $M$, the black hole charge $Q$, and the cosmological constant $\Lambda$, see Table \ref{TCoulomb1}. We can observe that the decay rate decreases, and the frequency of the oscillations increases, when the absolute value of the charge of the scalar fields increases. Also, the QNMs acquire a real part when scalar fields is charged. It is worth mentioning that for positive values of scalar field charge there is only a change of sign in the real part of the QNMs.\\

\begin {table}[H]
\caption {QNFs for massless charged scalar fields in the background  of three-dimensional Coulomb-like AdS black holes with $M=1$, $Q=0.10$, $\Lambda = -1 $, $\kappa=0$, and different values of the overtone number $n$, and $q$.}
\label {TCoulomb1}\centering
\scalebox{0.8}{
\begin {tabular} { | c | c | c |}
\hline
${}$ & $q = 0.00$ & $q= -0.01$   \\\hline
$\omega(n=0)$ &
$-1.83976265 i$ &
$-0.00623018 - 1.83966370 i$  \\\hline
$\omega(n=1)$ &
$-2.12181146 i$ &
$0.00409573 - 2.12191154 i$   \\\hline
$\omega(n=2)$ &
$-3.72425650 i$ &
$-0.00432750 - 3.72422698 i$  \\\hline
$\omega(n=3)$ &
$-4.18567498 i$ &
$0.00227199 - 4.18570504 i$  \\\hline
$\omega(n=4)$ &
$-5.63006461 i$ &
$-0.00352381 - 5.63004821 i$  \\\hline
${}$ & $q = -0.02$ & $q= -0.03$   \\\hline
$\omega(n=0)$ &
$-0.01243873 - 1.83936885 i$ &
$-0.01860501 - 1.83888393 i$  \\\hline
$\omega(n=1)$ &
$0.00816982 - 2.12220981 i$ &
$0.01220165 - 2.12270041 i$   \\\hline
$\omega(n=2)$ &
$-0.00865245 - 3.72413854 i$ &
$-0.01297234 - 3.72399144 i$  \\\hline
$\omega(n=3)$ &
$0.00454143 - 4.18579510 i$ &
$0.00680580 - 4.18594489 i$  \\\hline
$\omega(n=4)$ &
$-0.00704679 - 5.62999903 i$ &
$-0.01056811 - 5.62991713 i$  \\\hline
${}$ & $q = -0.04$ & $q= -0.05$   \\\hline
$\omega(n=0)$ &
$-0.02471026 - 1.83821811 i$ &
$-0.03073829 - 1.83738327 i$  \\\hline
$\omega(n=1)$ &
$0.01617247 - 2.12337418 i$ &
$0.02006608 - 2.12421926 i$   \\\hline
$\omega(n=2)$ &
$-0.01728469 - 3.72378616 i$ &
$-0.02158707 - 3.72352336 i$  \\\hline
$\omega(n=3)$ &
$0.00906261 - 4.18615393 i$ &
$0.01130945 - 4.18642157 i$  \\\hline
$\omega(n=4)$ &
$-0.01408694 - 5.62980261 i$ &
$-0.01760246 - 5.62965559 i$  \\\hline
${}$ & $q = -0.06$ & $q= -0.07$   \\\hline
$\omega(n=0)$ &
$-0.03667579 - 1.83639310 i$ &
$-0.04251248 - 1.83526239 i$  \\\hline
$\omega(n=1)$ &
$0.02386917 - 2.12522193 i$ &
$0.02757147 - 2.12636743 i$   \\\hline
$\omega(n=2)$ &
$-0.02587713 - 3.72320385 i$ &
$-0.03015261 - 3.72282862 i$  \\\hline
$\omega(n=3)$ &
$0.01354396 - 4.18674700 i$ &
$0.01576387 - 4.18712922 i$  \\\hline
$\omega(n=4)$ &
$-0.02111387 - 5.62947625 i$ &
$-0.02462037 - 5.62926480 i$  \\\hline
${}$ & $q = -0.08$ & $q= -0.09$   \\\hline
$\omega(n=0)$ &
$-0.04824100 - 1.83400623 i$ &
$-0.05385663 - 1.83263949 i$  \\\hline
$\omega(n=1)$ &
$0.03116564 - 2.12764064 i$ &
$0.03464693 - 2.12902670 i$   \\\hline
$\omega(n=2)$ &
$-0.03441138 - 3.72239879 i$ &
$-0.03865139 - 3.72191562 i.$  \\\hline
$\omega(n=3)$ &
$0.01796705 - 4.18756713 i$ &
$0.02015145 - 4.18805945 i$  \\\hline
$\omega(n=4)$ &
$-0.02812119 - 5.62902148 i$ &
$-0.03161555 - 5.62874655 i$  \\\hline
\end {tabular}}
\end{table}

{\bf{Case $\kappa\neq 0$.}} In order to show the behavior of the QNMs, we fix the black hole mass $M$, charge $Q$, and the cosmological constant $\Lambda$, see Table \ref{TCoulomb2}. Note that for uncharged scalar fields the QNFs are complex, but when
scalar field is charged appear two branches for the modes. The rate decay of the QNFs for both branches decreases and the oscillation of the frequencies increases for one branch and decreases for the other one when the absolute value of the charge of the scalar field increases.

\begin {table}[H]
\caption {QNFs for massless charged scalar fields in the background  of three-dimensional Coulomb-like AdS black holes with $M=1$, $Q=0.10$, $\Lambda = -1 $, $\kappa=1$, $n=0,1,2,3,4$, and different values of $q$. For positive values of scalar field charge there is only a change of sign in the real part of the QNMs.}
\label {TCoulomb2}\centering
\scalebox{0.8}{
\begin {tabular} { | c | c | c |}
\hline
${}$ & $q = 0.00$ & $q= -0.01$   \\\hline
$\omega(n=0)$ &
$0.98388504 - 1.98552182 i$ &
$-0.98470433 - 1.98445121 i$  \\\hline
${}$ & ${}$ &
$0.98306629 - 1.98659286 i$  \\\hline
$\omega(n=1)$ &
$0.95193811 - 3.96371161 i$ &
$-0.95284085 - 3.96224667 i$   \\\hline
${}$ & ${}$ &
$0.95103623 - 3.96517694 i$  \\\hline
$\omega(n=2)$ &
$0.91087011 - 5.93960613 i$ &
$-0.91181622 - 5.93787124 i$  \\\hline
${}$ & ${}$ &
$0.90992508 - 5.94134139 i$  \\\hline
$\omega(n=3)$ &
$0.86293045 - 7.91428723 i$ &
$-0.86390606 - 7.91233636 i$  \\\hline
${}$ & ${}$ &
$0.86195612 - 7.91623842 i$  \\\hline
$\omega(n=4)$ &
$0.80921835 - 9.88816723 i$ &
$-0.81021642 - 9.88602802 i$  \\\hline
${}$ & ${}$ &
$0.80822185 - 9.89030675 i$  \\\hline
${}$ & $q = -0.02$ & $q= -0.03$   \\\hline
$\omega(n=0)$ &
$-0.98552418 - 1.98338103 i$ &
$-0.98634457 - 1.98231128 i$  \\\hline
${}$ & $0.98224810 - 1.98766432 i$ & $0.98143045 - 1.98873621 i$ \\\hline
$\omega(n=1)$ &
$-0.95374444 - 3.96078213 i$ &
$-0.95464889 - 3.95931799 i$   \\\hline
${}$ & $0.95013521 - 3.96664266 i$ & $0.94923505 - 3.96810878 i$ \\\hline
$\omega(n=2)$ &
$-0.91276340 - 5.93613670 i$ &
$-0.91371164 - 5.93440252 i$  \\\hline
${}$ & $0.90898112 - 5.94307699 i$ & $0.90803823 - 5.94481295 i$ \\\hline
$\omega(n=3)$ &
$-0.86488296 - 7.91038584 i$ &
$-0.86586116 - 7.90843565 i$  \\\hline
${}$ & $0.86098309 - 7.91818993 i$ & $0.86001135 - 7.92014177 i$ \\\hline
$\omega(n=4)$ &
$-0.81121606 - 9.88388913 i$ &
$-0.81221726 - 9.88175056 i$  \\\hline
${}$ & $0.80722690 - 9.89244657 i$ & $0.80623353 - 9.89458669 i$ \\\hline
${}$ & $q = -0.04$ & $q= -0.05$   \\\hline
$\omega(n=0)$ &
$-0.98716551 - 1.98124195 i$ &
$-0.98798699 - 1.98017305 i$  \\\hline
${}$ & $0.98061336 - 1.98980852 i$ & $0.97979682 - 1.99088126 i$ \\\hline
$\omega(n=1)$ &
$-0.95555420 - 3.95785424 i$ &
$-0.95646036 - 3.95639090 i$   \\\hline
${}$ & $0.94833574 - 3.96957528 i $ & $0.94743730 - 3.97104216 i$ \\\hline
$\omega(n=2)$ &
$-0.91466096 - 5.93266870 i$ &
$-0.91561134 - 5.93093526 i$  \\\hline
${}$ & $0.90709641 - 5.94654925 i$ & $0.90615566 - 5.94828589 i$ \\\hline
$\omega(n=3)$ &
$-0.86684064 - 7.90648580 i$ &
$-0.86782142 - 7.90453630 i$  \\\hline
${}$ & $0.85904090 - 7.92209392 i$ & $0.85807174 - 7.92404637 i$ \\\hline
$\omega(n=4)$ &
$-0.81322002 - 9.87961232 i$ &
$-0.81422434 - 9.87747443 i$  \\\hline
${}$ & $0.80524171 - 9.89672709 i$ & $0.80425146 - 9.89886777 i$ \\\hline
${}$ & $q = -0.06$ & $q= -0.07$   \\\hline
$\omega(n=0)$ &
$-0.98880902 - 1.97910459 i$ &
$-0.98963160 - 1.97803656 i$  \\\hline
${}$ & $0.97898083 - 1.99195441 i $ & $0.97816539 - 1.99302799 i$ \\\hline
$\omega(n=1)$ &
$-0.95736737 - 3.95492796 i$ &
$-0.95827524 - 3.95346543 i$   \\\hline
${}$ & $0.94653971 - 3.97250943 i$ & $0.94564298 - 3.97397708 i$ \\\hline
$\omega(n=2)$ &
$-0.91656280 - 5.92920219 i$ &
$-0.91751532 - 5.92746949 i$  \\\hline
${}$ & $0.90521599 - 5.95002288 i$ & $0.90427738 - 5.95176020 i$ \\\hline
$\omega(n=3)$ &
$-0.86880348 - 7.90258716 i$ &
$-0.86978683 - 7.90063837 i$  \\\hline
${}$ & $0.85710388 - 7.92599914 i$ & $0.85613730 - 7.92795220 i$ \\\hline
$\omega(n=4)$ &
$-0.81523023 - 9.87533687 i$ &
$-0.81623767 - 9.87319968 i$  \\\hline
${}$ & $0.80326277 - 9.90100872 i$ & $0.80227565 - 9.90314993 i$ \\\hline
${}$ & $q = -0.08$ & $q= -0.09$   \\\hline
$\omega(n=0)$ &
$-0.99045472 - 1.97696896 i$ &
$-0.99127838 - 1.97590179 i$  \\\hline
${}$ & $0.97735051 - 1.99410198 i $ & $0.97653618 - 1.99517639 i$ \\\hline
$\omega(n=1)$ &
$-0.95918396 - 3.95200330 i$ &
$-0.96009354 - 3.95054159 i$   \\\hline
${}$ & $0.94474712 - 3.97544512 i $ & $0.94385211 - 3.97691352 i$ \\\hline
$\omega(n=2)$ &
$-0.91846891 - 5.92573716 i$ &
$-0.91942357 - 5.92400522 i$  \\\hline
${}$ & $0.90333986 - 5.95349785 i $ & $0.90240340 - 5.95523584 i$ \\\hline
$\omega(n=3)$ &
$-0.87077147 - 7.89868995 i$ &
$-0.87175740 - 7.89674189 i$  \\\hline
${}$ & $0.85517202 - 7.92990556 i $ & $0.85420803 - 7.93185921 i$ \\\hline
$\omega(n=4)$ &
$-0.81724668 - 9.87106284 i$ &
$-0.81825725 - 9.86892638 i$  \\\hline
${}$ & $0.80129009 - 9.90529140 i $ & $0.80030609 - 9.90743311 i$ \\\hline
\end {tabular}}
\end{table}

\begin{figure}[h]
\begin{center}
\includegraphics[width=0.45\textwidth]{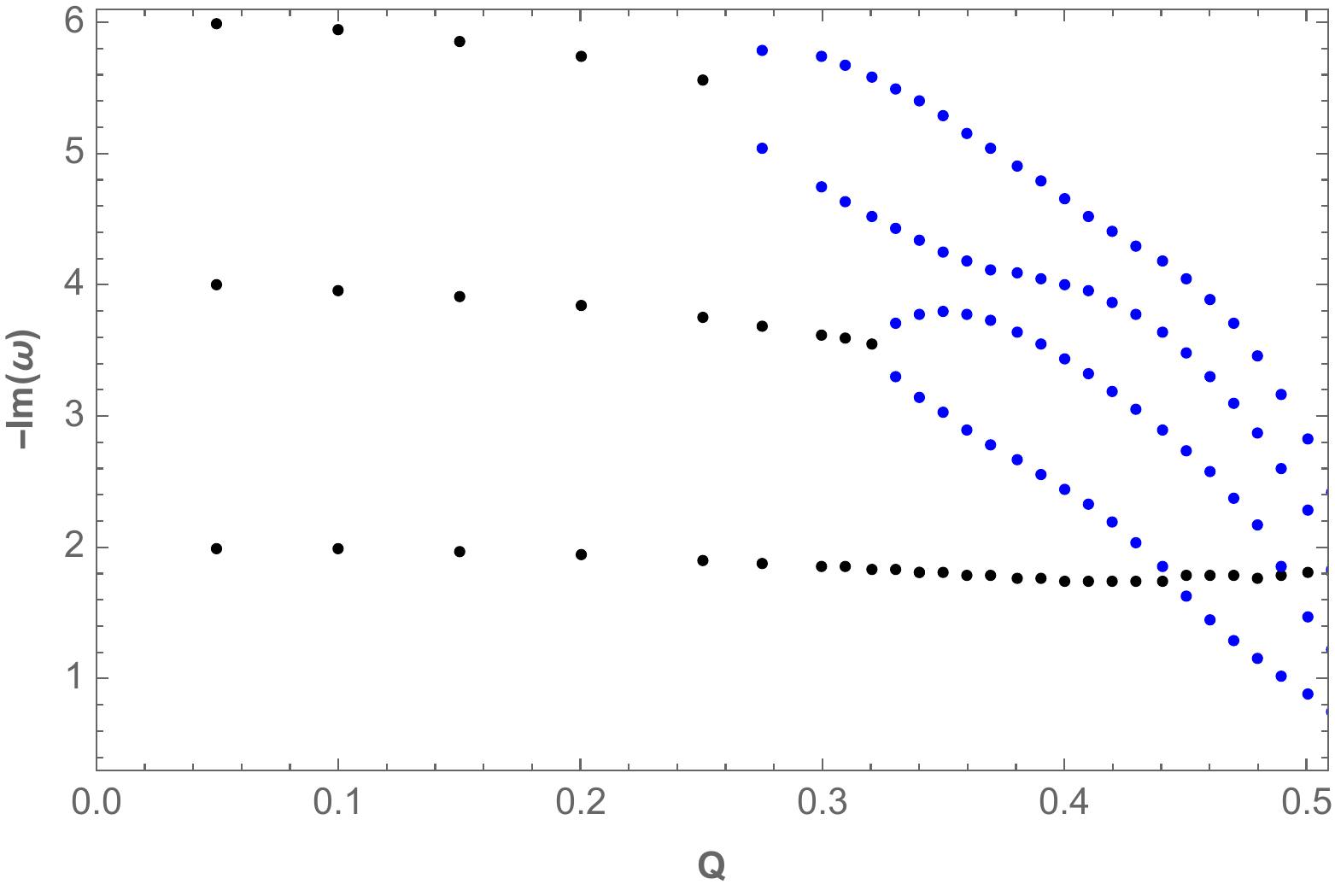}\\
\includegraphics[width=0.45\textwidth]{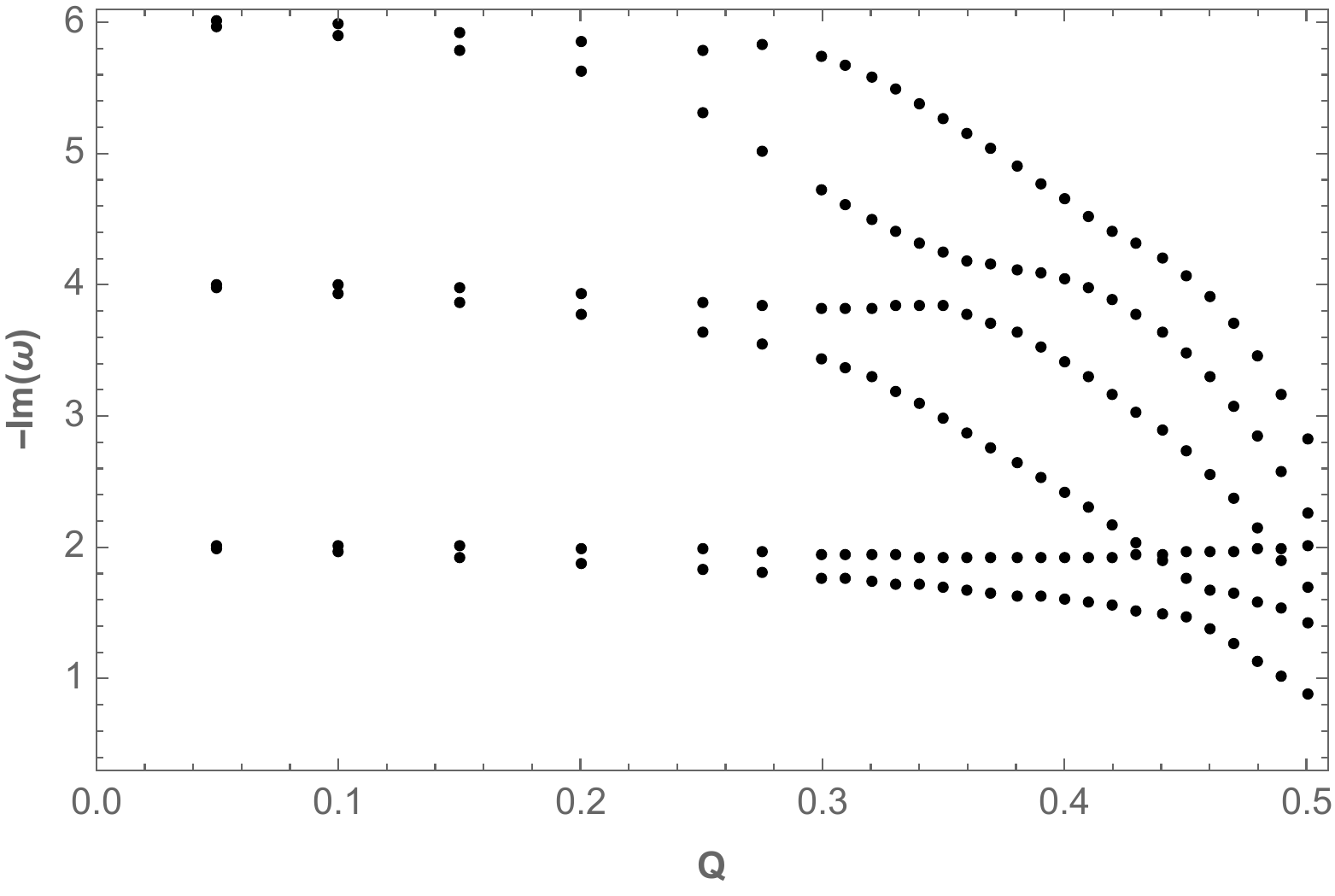}
\end{center}
\caption{QNFs for massless scalar fields in the background  of three-dimensional Coulomb like AdS black holes with $M=1$, $\Lambda=-1 $, $\kappa=1$, and different values of the overtone number $n$, and $q$. For the top panel, the black points correspond to complex QNFs, while that blue points correspond to purely imaginary QNFs; $n=0,1,2$, for small black hole charge and $n=0,1,2,3,4$, for bigger black hole charge \cite{Aragon:2021ogo}. Bottom panel corresponds to complex QNFs for charged scalar field $q= -0.25$, and $n=0,1,2,3,4,5$. For positive values of scalar field charge there is only a change of sign in the real part of the QNMs.}
\label{families}
\end{figure}

Also, for uncharged scalar fields, it was shown that there are complex QNFs for small values of the black hole charge that then become in two branches of imaginary QNFs when the black hole charge increases; thereby,  for small values of the black hole charge the complex QNFs are dominant, while that for bigger values of the black hole charge the purely imaginary QNFs are dominant. Also, the value of the charge for which occurs decreases when the overtone number increases, see top panel Fig. \ref{families}. However, this behavior disappears when the scalar field is charged, and the QNMs are always complex, see bottom panel Fig. \ref{families}.


\subsection{Massive charged scalar fields}

To analyze the propagation of massive charged scalar fields we show the behaviour of the QNFs for fixed values of black hole mass, cosmological constant, black hole charge, and a vanishing angular number, and different values of the overtone number and scalar field mass, see Table \ref{TCoulomb3}, where we can observe that when the scalar field mass increase the decay rate of the QNMs increases, and the frequency of the oscillations decreases. The same effect occurs for $\kappa>0$, see Table \ref{TCoulomb4}. On the other hand, note that the longest-lived modes are the ones with smallest angular number, contrary to Schwarzschild-AdS and Reissner-Nordstr\"om-AdS space-times where appear an anomalous behavior of the decay rate, i.e the longest-lived modes are the ones with higher angular number for small values of the scalar field mass \cite{Aragon:2020tvq,Fontana:2020syy}.

\begin {table}[H]
\caption {QNFs for massive scalar fields in the background of three-dimensional Coulomb-like AdS black holes with $M=1$, $\Lambda = -1 $, $\kappa=0$, and different values of the overtone number $n$, $Q=0.25$, and $q= -0.10$. For positive values of scalar field charge there is only a change of sign in the real part of the QNMs.}
\label {TCoulomb3}\centering
\scalebox{0.8}{
\begin {tabular} { | c | c | c |}
\hline
${}$ & $m = 0.00$ & $m=0.02$   \\\hline
$\omega(n=0)$ &
$-0.07307103 - 1.51833088 i$ &
$-0.07307092 - 1.51848883 i$  \\\hline
$\omega(n=1)$ &
$0.00967228 - 2.21275393 i$ &
$0.00966755 - 2.21295844 i$   \\\hline
$\omega(n=2)$ &
$-0.04307788 - 3.16477918 i$ &
$-0.04307604 - 3.16495013 i$  \\\hline
$\omega(n=3)$ &
$-0.02254329 - 4.15448750 i$ &
$-0.02254568 - 4.15467321 i$  \\\hline
${}$ & $m = 0.04$ & $m=0.06$   \\\hline
$\omega(n=0)$ &
$-0.07307058 - 1.51896252 i$ &
$-0.07307001 - 1.51975136 i$  \\\hline
$\omega(n=1)$ &
$0.00965334 - 2.21357170 i$ &
$0.00962969 - 2.21459294 i$   \\\hline
$\omega(n=2)$ &
$-0.04307053 - 3.16546280 i$ &
$-0.04306135 - 3.16631658 i$  \\\hline
$\omega(n=3)$ &
$-0.02255285 - 4.15523011 i$ &
$-0.02256478 - 4.15615752 i$  \\\hline
${}$ & $m = 0.08$ & $m=0.10$   \\\hline
$\omega(n=0)$ &
$-0.07306922 - 1.52085444 i$ &
$-0.07306821 - 1.52227044 i$  \\\hline
$\omega(n=1)$ &
$0.00959663 - 2.21602086 i$ &
$0.00955418 - 2.21785367 i$   \\\hline
$\omega(n=2)$ &
$-0.04304851 - 3.16751049 i$ &
$-0.04303204 - 3.16904318 i$  \\\hline
$\omega(n=3)$ &
$-0.02258147 - 4.15745427 i$ &
$-0.02260287 - 4.15911879 i$  \\\hline
${}$ & $m = 0.12$ & $m=0.14$   \\\hline
$\omega(n=0)$ &
$-0.07306697 - 1.52399771 i$ &
$-0.07306552 - 1.52603425 i$  \\\hline
$\omega(n=1)$ &
$0.00950241 - 2.22008908 i$ &
$0.00944139 - 2.22272431 i$   \\\hline
$\omega(n=2)$ &
$-0.04301194 - 3.17091288 i$ &
$-0.04298824 - 3.17311751 i$  \\\hline
$\omega(n=3)$ &
$-0.02262896 - 4.16114902 i$ &
$-0.02265969 - 4.16354251 i$  \\\hline
${}$ & $m = 0.16$ & $m=0.18$   \\\hline
$\omega(n=0)$ &
$-0.07306385 - 1.52837770 i$ &
$-0.07306196 - 1.53102541 i$  \\\hline
$\omega(n=1)$ &
$0.00937118 - 2.22575612 i$ &
$0.00929188 - 2.22918083 i$   \\\hline
$\omega(n=2)$ &
$-0.04296097 - 3.17565458 i$ &
$-0.04293015 - 3.17852130 i$  \\\hline
$\omega(n=3)$ &
$-0.02269501 - 4.16629637 i$ &
$-0.02273487 - 4.16940732 i$  \\\hline
${}$ & $m = 0.20$ & $m=0.30$  \\\hline
$\omega(n=0)$ &
$-0.07305986 - 1.53397438 i$ &
$-0.07304631 - 1.55311464 i$ \\\hline
$\omega(n=1)$ &
$0.00920358 - 2.23299433 i$ &
$0.00863105 - 2.25772262 i$ \\\hline
$\omega(n=2)$ &
$-0.04289582 - 3.18171453 i$  &
$-0.04267293 - 3.20244835 i$ \\\hline
$\omega(n=3)$ &
$-0.02277921 - 4.17287172 i$  &
$-0.02306542 - 4.19534346 i$ \\\hline
\end{tabular}}
\end{table}

\begin{table}[H]
\caption {The fundamental QNFs ($n=0$) for massive scalar fields in the background of three-dimensional Coulomb-like AdS black holes with $M=1$, $\Lambda = -1 $, and different values of $\kappa$, $Q=0.25$, and $q= -0.10$. For positive values of scalar field charge there is only a change of sign in the real part of the QNMs.}
\label {TCoulomb4}\centering
\scalebox{0.8}{
\begin {tabular} { | c | c | c |}
\hline
${}$ & $m = 0.00$ & $m=0.02$   \\\hline
$\omega(\k=1)$ &
$-0.91057032 - 1.87677763 i$ &
$-0.91055490 - 1.87696299 i$   \\\hline
$\omega(\k=10)$ &
$-9.97518437 - 1.96141343 i$ &
$-9.97518034 - 1.96160879 i$  \\\hline
$\omega(\k=30)$ &
$-29.9860098 - 1.9777749 i$ &
$-29.9860075 - 1.9779722 i$  \\\hline
${}$ & $m = 0.04$ & $m=0.06$   \\\hline
$\omega(\k=1)$ &
$-0.91050867 - 1.87751886 i$ &
$-0.91043165 - 1.87844455 i$   \\\hline
$\omega(\k=10)$ &
$-9.97516825 - 1.96219464 i$ &
$-9.97514811 - 1.96317027 i$  \\\hline
$\omega(\k=30)$ &
$-29.9860007 - 1.9785639 i$ &
$-29.9859894 - 1.9795492 i$  \\\hline
${}$ & $m = 0.08$ & $m=0.10$   \\\hline
$\omega(\k=1)$ &
$-0.91032391 - 1.87973895 i$ &
$-0.91018553 - 1.88140049 i $   \\\hline
$\omega(\k=10)$ &
$-9.97511994 - 1.96453452 i$ &
$-9.97508376 - 1.96628576 i$  \\\hline
$\omega(\k=30)$ &
$-29.9859736 - 1.9809271 i$ &
$-29.9859533 - 1.9826958 i$  \\\hline
${}$ & $m = 0.20$ & $m=0.30$   \\\hline
$\omega(\k=1)$ &
$-0.90903870 - 1.89513128 i$ &
$-0.90715135 - 1.91757604 i$   \\\hline
$\omega(\k=10)$ &
$-9.97478407 - 1.98075919 i$ &
$-9.97429137 - 2.00442337 i$  \\\hline
$\omega(\k=30)$ &
$-29.9857850 - 1.9973137 i$ &
$-29.9855084 - 2.0212150 i$  \\\hline
\end {tabular}}
\end{table}\leavevmode\newline

\subsection{Superradiant modes}

Now, we show in Table \ref{S1} the fundamental QNMs for massless charged scalar fields and in Table \ref{S2} for massive charge scalar fields for higher values of the scalar field charge. So, by comparing the real part of the dominant modes, with the superradiant condition of charged scalar fields (\ref{nbek}), we can see that all the unstable modes are superradiant. The value of the scalar field charge for which the modes became unstable decreases when the black hole charge increase. Also, its charge increases when the angular number increases.

\begin{widetext}

\begin{table}[H]
\caption {The fundamental QNFs ($n=0$) for massless scalar fields in the background of three-dimensional Coulomb-like AdS black holes with $M=1$, $\Lambda = -1 $, and different values of $\kappa$, $Q$ and $q$.}
\label {S1}\centering
\scalebox{0.8}{
\begin {tabular} { | c | c | c | c | c | c |}
\hline
$Q=0.35$ & $q = 1$ & $q = 5$  & $q = 7.5$ & $q = 10$ & $q = 15$ \\\hline
$\omega(\k=0)$ &
$0.696567171 - 1.092077211 i$ &
$2.41623216 - 0.45916273 i$ &
$3.25857846 - 0.21554737 i$ &
$4.01949101 - 0.05753746 i$ &
$5.31554772 + 0.06311880 i$   \\\hline
$\omega(\k=1)$ &
$1.13011346 - 1.40322475 i$ &
$2.62745730 - 0.56988551 i$ &
$3.43734843 - 0.28819352 i$ &
$4.18275252 - 0.10623460 i$ &
$5.47215492 + 0.03417677 i$  \\\hline
$\omega(\k=30)$ &
$30.0237557 - 1.9031054 i$ &
$30.2753038 - 1.6560555 i$ &
$30.4414571 - 1.5028306 i$ &
$30.6149234 - 1.3512196 i$ &
$30.9850685 - 1.0557237 i$  \\\hline

$Q=0.40$ & $q = 1$ & $q = 5$  & $q = 7.5$ & $q = 10$ & $q = 15$ \\\hline
$\omega(\k=0)$ &
$0.750214150 - 0.929501295 i$ &
$2.69125669 - 0.29604640 i$ &
$3.63949501 - 0.07840066 i$ &
$4.47440731 + 0.02825085 i$ &
$5.80780053 + 0.08115290 i$   \\\hline
$\omega(\k=1)$ &
$1.11636004 - 1.24969014 i$ &
$2.88350000 - 0.38975177 i$ &
$3.80765572 - 0.13373991 i$ &
$4.63618259 - 0.00568808 i$ &
$5.96267463 + 0.05123723 i$  \\\hline
$\omega(\k=30)$ &
$30.0206120 - 1.8829938 i$ &
$30.3103858 - 1.5981092 i$ &
$30.5036747 - 1.4216037 i$ &
$30.7070655 - 1.2474265 i$ &
$31.1462564 - 0.9109263 i$  \\\hline

$Q=0.45$ & $q = 1$ & $q = 5$  & $q = 7.5$ & $q = 10$ & $q = 15$ \\\hline
$\omega(\k=0)$ &
$0.803103196 - 0.732000639 i$ &
$2.98934522 - 0.13330507 i$ &
$4.03674155 + 0.01997223 i$ &
$4.89757158 + 0.06073762 i$ &
$6.25301017 + 0.09388473 i$    \\\hline
$\omega(\k=1)$ &
$1.08400715 - 1.04956819 i$ &
$3.16550253 - 0.20671137 i$ &
$4.20307608 - 0.01607947 i$ &
$5.06501810 + 0.03153813 i$ &
$6.40315505 + 0.06291190 i$   \\\hline
$\omega(\k=30)$ &
$30.0155012 - 1.8612511 i$ &
$30.3441327 - 1.5372586 i$ &
$30.5657150 - 1.3366546 i$ &
$30.8009505 - 1.1392668 i$ &
$31.3158585 - 0.7621306 i$   \\\hline

\end {tabular} 
}
\end{table}\leavevmode\newline

\begin{table}[H]
\caption {The fundamental QNFs ($n=0$) for massive scalar fields ($m=0.20$) in the background of three-dimensional Coulomb-like AdS black holes with $M=1$, $\Lambda = -1 $, and different values of $\kappa$, $Q$ and $q$.}
\label {S2}\centering
\scalebox{0.8} {

\begin {tabular} { | c | c | c | c | c | c |}
\hline
$Q=0.35$ & $q = 1$ & $q = 5$  & $q = 7.5$ & $q = 10$ & $q = 15$ \\\hline
$\omega(\k=0)$ &
$0.697887934 - 1.105664133 i$ &
$2.42394615 - 0.47034604 i$ &
$3.26948090 - 0.22442365 i$ &
$4.03380046 - 0.06377869 i$ &
$5.33899767 + 0.06069009 i$   \\\hline
$\omega(\k=1)$ &
$1.12921435 - 1.41772375 i$ &
$2.63411948 - 0.58096942 i$ &
$3.44720038 - 0.29703377 i$ &
$4.19585410 - 0.11245001 i$ &
$5.49429810 + 0.03239012 i$   \\\hline
$\omega(\k=30)$ &
$30.02358784 - 1.92208919 i$ &
$30.27647385 - 1.67364459 i$ &
$30.44347039 - 1.51954717 i$ &
$30.61778368 - 1.36705970 i$ &
$30.98963068 - 1.06978322 i$   \\\hline

$Q=0.40$ & $q = 1$ & $q = 5$  & $q = 7.5$ & $q = 10$ & $q = 15$ \\\hline
$\omega(\k=0)$ &
$0.751471569 - 0.941438377 i$ &
$2.699685536 - 0.305342406 i$ &
$3.651894098 - 0.084763746 i$ &
$4.491850543 + 0.024887234 i$ &
$5.835682721 + 0.078949416 i$   \\\hline
$\omega(\k=1)$ &
$1.11485439 - 1.26232271 i$ &
$2.890818033 - 0.399033555 i$ &
$3.818818134 - 0.140140488 i$ &
$4.652090593 - 0.008815955 i$ &
$5.989569880 + 0.049775423 i$   \\\hline
$\omega(\k=30)$ &
$30.0203274 - 1.9017788 i$ &
$30.31163137 - 1.61527481 i$ &
$30.50588628 - 1.43775358 i$ &
$30.71024836 - 1.26255206 i$ &
$31.15139472 - 0.92394879 i$  \\\hline

$Q=0.45$ & $q = 1$ & $q = 5$  & $q = 7.5$ & $q = 10$ & $q = 15$ \\\hline
$\omega(\k=0)$ &
$0.8042484137 - 0.7418322341 i$ &
$2.998611587 - 0.140101360 i$ &
$4.051755341 + 0.016762596 i$ &
$4.919499564 + 0.058792334 i$ &
$6.284376283 + 0.091577176 i$    \\\hline
$\omega(\k=1)$ &
$1.081800861 - 1.059436232 i$ &
$3.173497339 - 0.213624618 i$ &
$4.216404436 - 0.019182805 i$ &
$5.085588842 + 0.030367029 i$ &
$6.433681412 + 0.061307321 i$   \\\hline
$\omega(\k=30)$ &
$30.01507103 - 1.87981686 i$ &
$30.34542344 - 1.55396787 i$ &
$30.56809454 - 1.35219437 i$ &
$30.80442617 - 1.15362077 i$ &
$31.32154834 - 0.77401205 i$  \\\hline

\end {tabular}
}
\end{table}\leavevmode\newline

\end{widetext}

\clearpage

\section{Final remarks}
\label{conclusion}

In this work, we studied the propagation of charged scalar fields in the background of $2+1$-dimensional Coulomb-like AdS black holes, and we showed that such propagation is not always stable under Dirichlet boundary conditions. Then, we solved the Klein-Gordon equation by using the pseudospectral Chebyshev method, and we found the corresponding quasinormal frequencies.  Mainly, we showed that, when the scalar field is charged, the QNFs are always complex, contrary to the uncharged case, where for small values of the black hole charge, the complex QNFs are dominant, while that for bigger values of the black hole charge the purely imaginary QNFs are dominant. Last by not least, we found that all the unstable modes are superradiant and all the stable modes are not superradiant, according to the superradiant condition, and consequently, the scalar waves can  experiment a superradiant amplification  by the black hole by extracting charged of the black hole indicating that the black hole geometry is unstable.\\

\acknowledgments
This work is partially supported by ANID Chile through FONDECYT Grant No 1170279 (J. S.). The author A. R. acknowledges Universidad de Tarapac\'a for financial support.

\newpage


\end{document}